%% file: neurips_2024.tex
\documentclass{article}

\usepackage[final]{neurips_2024}
\usepackage{graphicx}
\usepackage{subfigure}
\usepackage{booktabs} 
\usepackage{enumitem}
\usepackage{multicol}
\usepackage{pythonhighlight}

\usepackage[utf8]{inputenc} 
\usepackage[T1]{fontenc}    
\usepackage{hyperref}       
\usepackage{url}            
\usepackage{booktabs}       
\usepackage{amsfonts}       
\usepackage{nicefrac}       
\usepackage{microtype}      
\usepackage{xcolor}         
\usepackage{multirow}

\usepackage{amsmath}
\usepackage{amssymb}
\usepackage{mathtools}
\usepackage{amsthm}
\usepackage{caption}
\usepackage{tcolorbox}
\usepackage{booktabs}
\usepackage{arydshln}
\usepackage[textsize=tiny]{todonotes}
\newtcolorbox{mybox}{colframe = black!10!black}

\usepackage[capitalize,noabbrev]{cleveref}

\theoremstyle{plain}

\theoremstyle{definition}

\theoremstyle{remark}

\title{UniMixer: A Unified Architecture for Scaling Laws in Recommendation Systems}

\author{
Mingming Ha \quad
Guanchen Wang \quad
Linxun Chen \quad
Xuan Rao \quad
Yuexin Shi \quad
Tianbao Ma \\
\textbf{Zhaojie Liu \quad
Yunqian Fan\quad
Zilong Lu\quad
Yanan Niu\quad
Han Li \quad
Kun Gai} \\[2pt]
Kuaishou Technology, Beijing, China \\[2pt]
\{hamingming, wangguanchen, chenxi36, raoxuan, shiyuexin, matianbao, \\zhaotianxing, fanyunqian03, luzilong, niuyanan, lihan08\}@kuaishou.com,\;gaikun@qq.com
}

\begin{document}

\maketitle

\begin{abstract}
In recent years, the scaling laws of recommendation models have attracted increasing attention, which govern the relationship between performance and parameters/FLOPs of recommenders. Currently, there are three mainstream architectures for achieving scaling in recommendation models, namely attention-based, TokenMixer-based, and factorization-machine-based methods, which exhibit fundamental differences in both design philosophy and architectural structure. In this paper, we propose a unified scaling architecture for recommendation systems, namely \textbf{UniMixer}, to improve scaling efficiency and establish a unified theoretical framework that unifies the mainstream scaling blocks. By transforming the rule-based TokenMixer to an equivalent parameterized structure, we construct a generalized parameterized feature mixing module that allows the token mixing patterns to be optimized and learned during model training. Meanwhile, the generalized parameterized token mixing removes the constraint in TokenMixer that requires the number of heads to be equal to the number of tokens. Furthermore, we establish a unified scaling module design framework for recommender systems, which bridges the connections among attention-based, TokenMixer-based, and factorization-machine-based methods. To further boost scaling ROI, a lightweight UniMixing module is designed, \textbf{UniMixing-Lite}, which further compresses the model parameters and computational cost while significantly improve the model performance. The scaling curves are shown in the following figure. Extensive offline and online experiments are conducted to verify the superior scaling abilities of \textbf{UniMixer}.
\end{abstract}
\begin{figure}[ht]
\begin{center}
\centerline{\includegraphics[width=0.7\columnwidth]{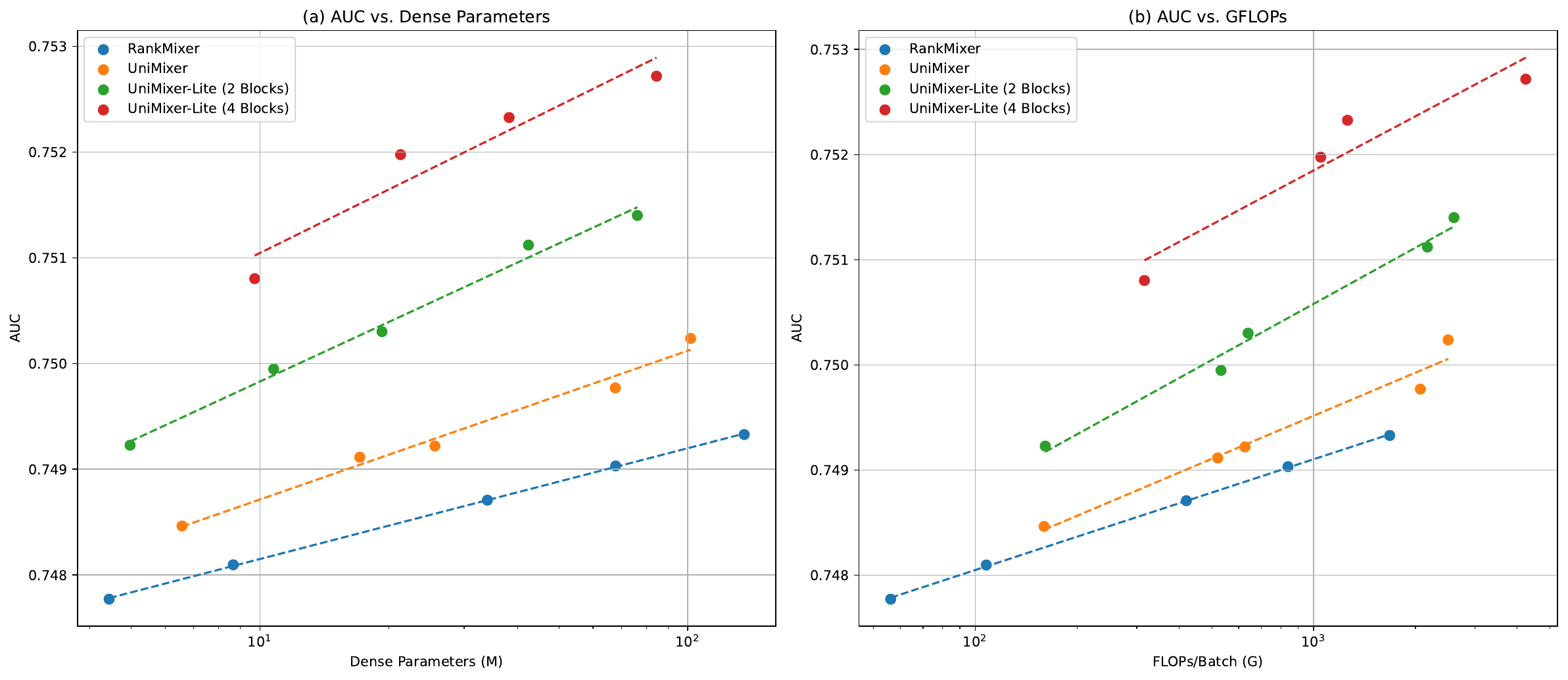}}
\caption{The scaling laws between AUC and Parameters for the present UniMixer/UniMixer-Lite and RankMixer architectures. The x-axis is presented on a logarithmic scale.}
\label{ScalingCurves_1}
\end{center}
\end{figure}

\input{sections/Introduction}
\input{sections/RelatedWorks}
\input{sections/Preliminaries}

\input{sections/UniMixer}
\input{sections/Experiments}
\input{sections/Conclusions}



\bibliographystyle{unsrt}
\bibstyle{neurips_2024:}
\bibliography{references}

\appendix
\include{sections/Appendix}

\newpage

\end{document}

%% file: sections/Introduction.tex
\section{Introduction}
\label{Introduction}
Large language models (LLMs) have revealed an impressive phenomenon: as model size, data volume, and computational resources increase, performance improves steadily, namely scaling laws. The remarkable performance scaling observed in LLMs has inspired the recommender systems community to explore scaling frameworks tailored to recommendation tasks. In recent years, researchers have attempted to design scaling modules and stack them across multiple layers to increase the complexity of ranking models, thereby achieving scaling laws between model performance and model size or computational cost (e.g., parameters and FLOPs).

Based on a large amount of multi-field user and item feature, recommender systems predict the user behaviors to present the most relevant content to them to increase user’s positive engagements with the recommendations. These multi-field features generally involve categorical features and dense features, which generally possess more dynamic embedding representations and capture information from multiple perspectives. Differing from the natural language processing (NLP) domain, where all tokens share a unified embedding space, the feature space in recommendation tasks is inherently heterogeneous. Therefore, learning heterogeneous features interactions represents a fundamental difference from NLP domain. Owing to the tremendous success of Transformers in LLMs, a natural idea is to modify the Transformer module to adapt recommendation tasks because it is generally infeasible to directly adopt Transformer module as the fundamental block for scaling laws in recommendation systems. To address the heterogeneous feature interaction problem, current mainstream scaling architectures for recommendation models can be broadly categorized into three types: attention-based, TokenMixer-based, and factorization-machine-based methods. Attention-based methods (e.g., HiFormer \cite{gui2023hiformer}, FAT \cite{yan2025scaling}, and HHFT \cite{yu2025hhft}, etc.) construct token-specific query, key, and value projections for each input token. In contrast to attention-based methods, TokenMixer-based methods (e.g., RankMixer \cite{zhu2025rankmixer}, TokenMixer-Large \cite{jiang2026tokenmixer}, etc.) employ the rule-based token mixing operation to achieve heterogeneous feature interactions, which avoids computing inner-product similarity between two heterogeneous semantic spaces. Factorization-machine-based methods (e.g. Wukong \cite{zhang2024wukong}, Kunlun \cite{hou2026kunlun}, etc.), on the other hand, model feature interactions by introducing a Factorization Machine (FM) block to compute interactions among the input embeddings within each layer. These frameworks are built upon completely different scaling blocks, yet all demonstrate the capability to scale up model performance. This leads us to a fundamental question: Can we construct a unified scaling module for recommendation systems that combines the advantages of existing mainstream scaling components? To bridge the connections among these scaling modules, we first find a parameterized formulation of the rule-based TokenMixer operation. By further optimizing the computation pipeline, we derive the UniMixing module with reduced computational cost. Based on this design and results, we propose a unified theoretical framework that unifies the mainstream scaling modules in recommender systems. Besides, a lightweight UniMixer module is designed, which use the advantages of existing mainstream scaling blocks and achieve the best parameter efficiency and computational efficiency. We hope that the unified architecture can help the recommendation systems community achieve its own ``attention moment''. Our main contributions can be summarized as follows: 
\begin{itemize}
    \item[1)] We reveal the feature interaction patterns of TokenMixer via equivalent parameterization of the rule-based TokenMixer.
    \item[2)] We propose a unified scaling framework, termed UniMixer, which bridges the differences and connections among attention-based, TokenMixer-based, and FM-based methods. By optimizing the computation pipeline, UniMixer significantly reduces computational complexity and GPU memory consumption during both training and inference.
    \item[3)] To further reduce model parameters and computational cost, we design a lightweight UniMixing module, called UniMixing-Lite, which can simultaneously leverage the advantages of both attention-based and TokenMixer-based architecture to achieve improved scaling efficiency.
    \item[4)] Extensive offline and online experiments are performed to demonstrate the superior scaling abilities of UniMixer.
\end{itemize}

%% file: sections/RelatedWorks.tex
\section{Related Work} 
Currently, there are three scaling modeling paradigms for establishing 
scaling laws for massive-scale recommendation systems: attention-based, TokenMixer-based,and FM-based methods.
\paragraph{Attention-Based Framework.} 
Recent research in recommendation systems has adapted Transformers for CTR prediction. A core challenge in this paradigm is bridging the gap between the heterogeneous nature of the token sequence and the sequential compositionality that assumed by language modeling. To this end, in \cite{gui2023hiformer}, the heterogeneous attention layer is proposed to address the heterogeneous feature interaction and HiFormer is designed to explicitly model high-order interactions by flattening heterogeneous tokens into a single vector representation. Additionally, Field-Aware Transformers (FAT) inject field-aware interaction priors into the attention mechanism via factorized contextual alignment and cross-field modulation \cite{yan2025scaling}, further establishing the empirical scaling law for CTR prediction. HHFT further validates these scaling properties by interleaving heterogeneous Transformer blocks (for preserving domain-specific semantics) with HiFormer blocks (for high-order interaction learning) \cite{yu2025hhft}. Furthermore, in dynamic user behavior modeling, methods such as HSTU-V1/V2 \cite{10.5555/3692070.3694484, ding2026bending}, MARM \cite{10.1145/3746252.3761015}, OneTrans \cite{zhang2025onetrans}, Climber \cite{10.1145/3746252.3761561}, Hyformer \cite{huang2026hyformer}, and LLaTTE \cite{xiong2026llatte} leverage attention mechanisms to capture long-range temporal dependencies. These approaches highlight the potential of unifying feature interaction and sequential behavior modeling to achieve more robust scaling laws.

\paragraph{TokenMixer-Based Framework.}
While attention mechanisms offer expressive feature interaction, they incur prohibitive computational costs due to the quadratic complexity of attention score computation. Inspired by the success of MLP-Mixer \cite{tolstikhin2021mlp} in computer vision, a paradigm shift towards token-mixing architectures has emerged in industrial recommender systems, yielding advanced models such as RankMixer \cite{zhu2025rankmixer}, Lemur \cite{han2025lemur}, and TokenMixer-Large \cite{jiang2026tokenmixer}. For instance, RankMixer replaces dynamic attention with static, non-parametric token-mixing operations, achieving competitive CTR prediction performance while maintaining strictly comparable FLOPs \cite{zhu2025rankmixer}. Building upon this, TokenMixer-Large scales this architecture to 13B configurations by introducing auxiliary residual connections and tailored loss functions \cite{jiang2026tokenmixer}, demonstrating compelling scaling laws across various model dimensions. Nonetheless, a critical gap remains: the design of current token-mixing operators heavily relies on empirical rules and lacks a rigorous theoretical bridge to traditional FM-based or attention-based methodologies.

\paragraph{FM-Based Framework.}
The pioneer FM-based method employs the low-order pairwise modeling for feature interactions in recommendation systems \cite{rendle2010factorization}, which was subsequently generalized by Field-aware FMs to capture field-specific and context-sensitive interactions \cite{juan2016field}. While these models benefit from high interpretability and efficiency, they are constrained intrinsically by their capacity of low-order interactions. To address this limitation, various neural network-based extentions, such as DeepFM \cite{guo2017deepfm}, AutoInt \cite{song2019autoint}, and DCN series \cite{wang2017deep,wang2021dcnv2}, integrate MLP or transformer attention to capture high-order interactions. More recently, Wukong \cite{zhang2024wukong} has demonstrated appropriate scaling properties by stacking FM-style interaction blocks with linear compression. Nevertheless, the reliance on explicit low-order interaction of FM-based methods still limits the performance improvement when models are scaled up in terms of parameters and FLOPs, which is in contrast to the predictive scaling laws observed in LLMs \cite{kaplan2020scaling, 10.5555/3600270.3602446}.

%% file: sections/Preliminaries.tex
\section{Preliminaries}
\label{Preliminaries}
Consider a class of discriminative recommendation tasks, such as rating, click-through rate (CTR) and post-click conversion rate (CVR) predictions, and so forth, which are typically formulated as a supervised learning problem. The dataset is defined as 
$\mathcal{D}=\{(\textbf{X}_1, y_1),\dotsc,(\textbf{X}_i, y_i),\dotsc,(\textbf{X}_N, y_N)\}$, where $\textbf{X}_i=\big[\textbf{x}_i^{(1)},\textbf{x}_i^{(2)},\dotsc,\textbf{x}_i^{(F)}\big]$ with $F$ feature fields, $y_i\in\{0,1\}$ corresponding to a binary classification problem or $y_i\in\mathbb{R}$ for a regression problem is the label of the $i$-th sample, $N$ is the number of data points. In general, the input features $\textbf{X}=\{\textbf{X}^\text{C},\textbf{X}^\text{D}\}$ are divided into categorical features $\textbf{X}^\text{C}$ and dense features $\textbf{X}^\text{D}$. $\vert{C}\vert$ and $\vert{D}\vert$ are used to denote the numbers of categorical and dense features, respectively. For CTR and CVR prediction tasks, the core objective is to establish a model to predict the click or conversion probability $\text{Pr}(y_i=1\vert\textbf{X}_i)$. In recommender systems, the learned embedding representations are more dynamic \cite{gui2023hiformer}. Differing from input tokens of language models, the feature spaces are inherently heterogeneous \cite{zhu2025rankmixer}. Therefore, it is inappropriate to directly transfer the Transformer architecture used in large language models to recommendation modeling. To date, scaling laws in the recommendation domain have primarily been established through three types of foundational blocks and their variants. 
\paragraph{Heterogeneous Attention Layer.} Heterogeneous-attention-based architecture \cite{gui2023hiformer,yan2025scaling,yu2025hhft} generally use the field-specific query, key, and value projections to achieve heterogeneous feature interaction. Given an input hidden states ${X}=[\boldsymbol{x}_1;\dotsc;\boldsymbol{x}_T]\in\mathbb{R}^{T\times{D}}$, the heterogeneous attention layer is formulated as 
\begin{equation}
\label{Eq3_01}
{Q}_h=\left[
\begin{array}{c}
\boldsymbol{x}_1W^{1h}_Q\\ 
\vdots\\ 
\boldsymbol{x}_TW^{Th}_Q \\ 
\end{array}
\right]\in\mathbb{R}^{T\times{d}},
{K}_h=\left[
\begin{array}{c}
\boldsymbol{x}_1W^{1h}_K\\ 
\vdots\\ 
\boldsymbol{x}_TW^{Th}_K \\ 
\end{array}
\right]\in\mathbb{R}^{T\times{d}},
{V}_h=\left[
\begin{array}{c}
\boldsymbol{x}_1W^{1h}_V\\ 
\vdots\\ 
\boldsymbol{x}_TW^{Th}_V \\ 
\end{array}
\right]\in\mathbb{R}^{T\times{d}},
\end{equation}
where $W^{ih}_Q, W^{ih}_K, W^{ih}_V\in\mathbb{R}^{D\times{d}}$ are the token-specific weights of query, key, and value projections, respectively. 
The output of the multi-head heterogeneous attention layer is computed as follows
\begin{equation}
{O}_h=
\text{softmax}\Big(\frac{{Q}_h{K}_h^{\mathsf{T}}}{\sqrt{d}}\Big){V}_h\in\mathbb{R}^{T\times{d}}.
\end{equation}
Then the outputs of the multi-head heterogeneous attention are concatenated and passed through a linear projection to align the output dimension with the input ${X}$.
\paragraph{TokenMixer.} TokenMixer-based framework \cite{zhu2025rankmixer,jiang2026tokenmixer,qi2025mtmixattintegratingmixtureofexpertsmultimix} employ the parameter-free and rule-based mixing operation to perform the feature interaction. For the given input ${X}=[\boldsymbol{x}_1;\dotsc;\boldsymbol{x}_T]$, TokenMixer first evenly splits each input token $\boldsymbol{x}_t$ into $H$ heads.
\begin{equation}
\left[
\boldsymbol{x}_t^{(1)}\vert\;\boldsymbol{x}_t^{(2)}\;\vert\;\dotsc\;\vert\;\boldsymbol{x}_t^{(H)}
\right]
=\text{SplitHead}(\boldsymbol{x}_{t}).
\end{equation}
Then, the $h$-token $\boldsymbol{s}^h$ can be obtained as
\begin{equation}
\boldsymbol{s}^h=\text{concat}\big(\boldsymbol{x}_1^{(h)},\boldsymbol{x}_2^{(h)},\dotsc,\boldsymbol{x}_T^{(h)}\big)\in\mathbb{R}^{\frac{TD}{H}}
\end{equation}
The output of TokenMixer is formulated as
\begin{equation}
{S}=
\left[
\begin{array}{c}
\boldsymbol{s}_1\\ 
\vdots\\ 
\boldsymbol{s}_H\\ 
\end{array}
\right]\in\mathbb{R}^{H\times{\frac{TD}{H}}},
\end{equation}
where $H$ is required to be the same as $T$. Therefore, dimensions of the input ${X}$ and the output ${S}$ are identical.

\paragraph{Wukong.}
Wukong-based models \cite{zhang2024wukong,hou2026kunlun}concatenate the outputs of a Factorization Machine Block (FMB) and a linear projection layer to upscale the interaction component. 
\begin{equation}
    \label{wukong}
\begin{aligned}
&\text{FMB}(X)=\;\text{reshape}(\text{MLP}(\text{LN}(\text{flatten}(\text{FM}(X))))),\;\text{FM}(X)=XX^{\mathsf{T}}Y\\
&\text{LCB}(X)=\;WX
\end{aligned}
\end{equation}
where $W\in\mathbb{R}^{n\times{T}}$ and $Y\in\mathbb{R}^{T\times{r}}$ are learnable projection matrix. $Y$ is used to reduce memory requirement to store the interaction matrix $XX^{\mathsf{T}}$.

In this work, we focus on establishing a unified structural foundation for recommendation systems that integrates the strengths of current scaling blocks to further increase the scaling ROI. 


%% file: sections/UniMixer.tex
\section{UniMixer}
\label{UniMixer}
\subsection{Overview}
\begin{figure}[ht]
\centering
\centerline{\includegraphics[width=0.9\columnwidth]{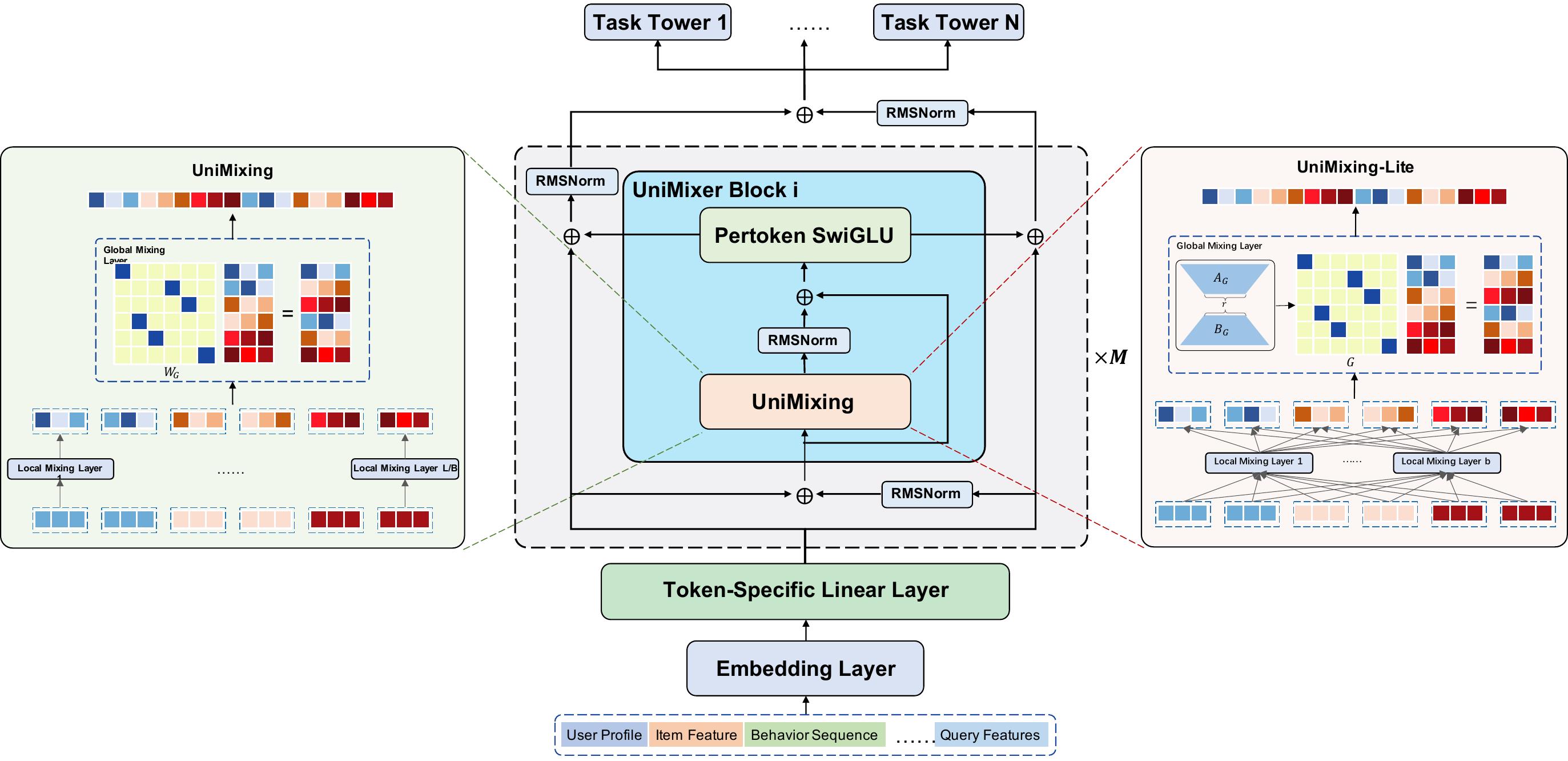}}
\caption{The UniMixer architecture for scaling laws in recommendation systems.}
\label{framework}
\end{figure}

A unified module, namely the UniMixer block, for scaling up recommender systems is established, which unifies the mainstream scaling modules for recommendation such as attention-based modules, TokenMixer-based modules, and Wukong-based methods, under a unified theoretical framework. As shown in Fig. \ref{framework}, the overall architecture consists of feature tokenization, $M$ UniMixer blocks with the Siamese norm and Sparse-Pertoken MoE. Through parameterized rule-based TokenMixer, we bridge the connection among attention-based, TokenMixer-based, and Wokong-based methods, enabling the proposed UniMixer to simultaneously possess the advantages of these approaches. Besides, a lightweight UniMixing module is developed to further compresses the
model parameters and computational cost while significantly improve the model performance.
\subsection{Feature Tokenization}
Based on the semantic categories of the input feature fields, the input features $\textbf{X}$ is first divided into $N$ disjoint feature domains
\begin{equation}
\textbf{X} = \Big[
\underset{\text{User Profile}}{\underbrace{\textbf{x}_U^{(1)},\dotsc, \textbf{x}_U^{(n_U)}}},
\underset{\text{Item Features}}{\underbrace{\textbf{x}_I^{(1)},\dotsc, \textbf{x}_I^{(n_I)}}},
\underset{\text{Behavior Sequence}}{\underbrace{\textbf{x}_B^{(1)},\dotsc, \textbf{x}_B^{(n_B)}}},
\underset{\text{Query Features}}{\underbrace{\textbf{x}_Q^{(1)},\dotsc, \textbf{x}_Q^{(n_Q)}}},
\dotsc
\Big].
\end{equation}
Each feature domain is transformed into different embedding vectors with dimension by embedding layers
\begin{equation}
\textbf{e}_n = \text{Embedding}(\textbf{X}_\text{domain})\in\mathbb{R}^{d_\text{domain}},
\end{equation}
where $\textbf{X}_\text{domain}$ denotes all the features within a feature domain, $d_\text{domain}$ is the embedding dimension corresponding this feature domain. The obtained feature domain embeddings are concatenated into one embedding vector $\textbf{E}=[\textbf{e}_1,\textbf{e}_2,\dotsc,\textbf{e}_N]$. Similar to \cite{zhu2025rankmixer}, the embedding vector $\textbf{E}$ is evenly divided into an appropriate number of block. Then, each block is projected into a token embedding by using the following token-specific linear layer
\begin{equation}
    \boldsymbol{x}_i=W^\text{proj}_i\textbf{E}_{di:di+d}+\textbf{b}^\text{proj}_i\in\mathbb{R}^{D},
\end{equation}
where $W^\text{proj}_i\in\mathbb{R}^{D\times{d}}$, $\textbf{b}^\text{proj}_i\in\mathbb{R}^{D}$. The input hidden states $X\in\mathbb{R}^{T\times{D}}$ can then be obtained by stacking $\boldsymbol{x}_i$ column-wise.

\subsection{UniMixer Block}
\begin{figure}[ht]
\centering
\begin{minipage}{0.45\columnwidth}
    \centering
    \includegraphics[width=\linewidth]{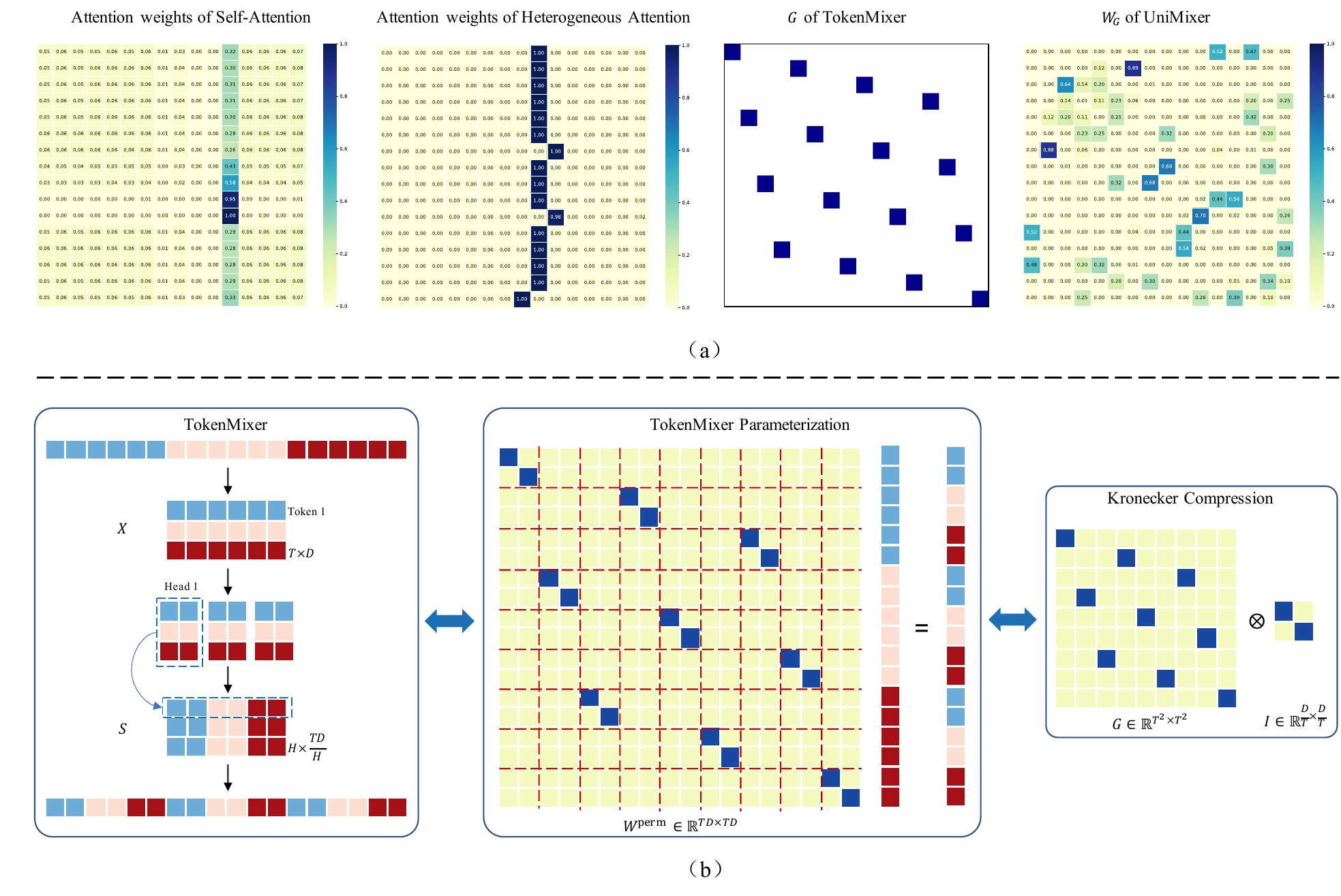}
    \caption{(a) The global mixing weights of different methods. (b) Equivalent parameterization of the rule-based TokenMixer.}
    \label{perm_weight}
\end{minipage}
\hfill
\begin{minipage}{0.5\columnwidth}
    As illustrated in the Fig. \ref{perm_weight}(a), it can be observed that the attention weights of the heterogeneous attention are sharp and sparse, which pose a risk to gradient backpropagation, thereby making the training of the query and key weights difficult and potentially causing it to stall, as shown in Fig. \ref{perm_weight}(a) (the $10$-th and $15$-th row in attention weights of the heterogeneous attention). Meanwhile, under large-scale heterogeneous feature inputs, such attention patterns may lead to uniform feature interactions, namely, attention scores become very small and lack discriminability, which potentially result in noise signals to obscure critical feature interaction patterns. 
\end{minipage}
\end{figure}

\paragraph{Heterogeneous Feature Interactions.} As mentioned in Section \ref{Preliminaries}, heterogeneous attention addresses the feature interaction problem between two heterogeneous semantic spaces by employing token-specific query, key, and value weights. However, the attention pattern obtained by computing the inner-product similarity typically carries a diagonally dominant prior. In the early stage of training, with randomly initialized weights $W_Q^h$ and $W_K^h$, the magnitude of the attention weights is largely dominated by the input token values $X$, which can easily cause the attention weights to concentrate on a small number of tokens, as shown in Fig. \ref{perm_weight}(a). 

On the other hand, the parameter-free and rule-based TokenMixer operation lacks learnability and scenario adaptability, which can lead to insufficient or erroneous heterogeneous feature interactions. In addition, requiring $T=H$ further restricts the selection of heterogeneous feature interaction patterns. Through in-depth analysis of the TokenMixer operation, we have made some interesting findings, which make it possible to parameterize the TokenMixer operation. As shown in Figure \ref{perm_weight}(b), we observe that the TokenMixer operation can be regarded as the product of a permutation matrix $W^\text{perm}$ and the flattened input embedding $\text{flatten}(X)\in\mathbb{R}^{TD}$, which can be formulated as 
\begin{equation}
\label{Eq4_3_1}
\text{TokenMixer}(X)=\text{reshape}(W^\text{perm}\text{flatten}(X)),
\end{equation}
where $W^\text{perm}\in\mathbb{R}^{TD\times{TD}}$ is a large permutation matrix. A concrete numerical example is provided in Appendix \ref{NumericalExample}. A natural idea is to enable rule-based TokenMixer to be learnable and optimizable by parameterizing the permutation matrix $W^\text{perm}$. However, the computation complexity $O(T^2D^2)$ and the number of parameters $O(T^2D^2)$ is unacceptable. Through observation, we have made some interesting findings regarding the permutation matrix $W^\text{perm}$ of TokenMixer and summarize them as the insightful properties in the following box.

\begin{mybox}
\textit{\textbf{Properties of The Permutation Matrix $W^\text{perm}$.}}
\begin{itemize}[leftmargin=*]
\item \textbf{Compressibility.} The permutation matrix $W^\text{perm}$ can be equivalently decomposed into the Kronecker product of two smaller matrices, i.e.,
 $W^\text{perm}=G\otimes{I}$, where $I\in\mathbb{R}^{\frac{D}{T}\times\frac{D}{T}}$ is the identity matrix and $G\in\mathbb{R}^{T^2\times{T}^2}$, $\otimes$ is the Kronecker product operation.
\item \textbf{Doubly Stochasticity.} In the permutation matrix $W^\text{perm}$, every row and every column sums to 1, i.e., $\sum_{p=1}^{TD}w^\text{perm}_{pq}=1, \sum_{q=1}^{TD}w^\text{perm}_{pq}=1$.
\item \textbf{Sparsity.} Each row/column of $W^\text{perm}$ contains exactly one non-zero element.
\item \textbf{Symmetry.} If $T=H$, the permutation matrix $W^\text{perm}$ is a symmetric matrix, i.e.,
 $W^\text{perm}=W^{\text{perm}\mathsf{T}}$. If $T\ne{H}$, the permutation matrix $W^\text{perm}$ is a asymmetric matrix.
\end{itemize}
\end{mybox}

According to the properties of the permutation matrix of TokenMixer, the number of parameters for token mixing is significantly reduced by parameterizing the matrices $G$ and $I$, namely, $O(T^4+(\frac{D}{T})^2)$, where $T$ is typically much smaller than $D$. Besides, there remain three challenges in the parameterization of the TokenMixer: (1) Directly using parameterized $G$ and $I$ to reconstruct $W^\text{perm}$ still produces an intermediate variable of size $[TD, TD]$ during model training and inferring processes, which imposes a very high demand on GPU memory; (2) How to ensure that the learned parameters satisfy doubly stochasticity, sparsity and symmetry; (3) How to design a unified recommendation scaling module that integrates the strengths of existing scaling modules and establish superior scaling efficiency for the recommendation systems.

\paragraph{Unified Token Mixing Module.} Inspired by Fig. \ref{perm_weight}, in the unified token mixing module, $T$ and $D$ are no longer used; instead, we define the block and the block size in the permutation matrix. The block size is denoted as $B$. The number of blocks is $(L//B)^2$, where $L$ is the input embedding dimension and can be divisible by the block size $B$. Denote the parameterized weights of $G$ as $W_G$. Considering the sparsity of the permutation matrix and to achieve sufficient heterogeneous feature interactions, we assign a distinct parameterized weight $W^i_B$ to each row. With this operation, each block possesses a different feature interaction pattern.
Then, a permutation matrix $W^\text{perm}$ with richer interaction patterns can be obtained by learning the parameter matrices $W_G$ and $W^i_B$, which is formulated as
\begin{equation}
\label{Eq4_3_2}
\text{UniMixing}(X)=\text{reshape}\Big(\Big(W_G\otimes\{W^i_B\}_{i=1}^{L//B}\Big)\text{flatten}(X),1,L\Big),
\end{equation}
where $\otimes$ is the generalized Kronecker product.

Next, the computation pipeline of (\ref{Eq4_3_2}) is optimized to  significantly reduce both the computational cost and GPU memory requirements. 
The embedding vector $\text{flatten}(X)$ is first evenly split into $L//B$ vectors and the size of each vector is $B$, which is expressed as 
\begin{equation}
\label{Eq4_3_3}
\left[
\boldsymbol{x}_1\vert\;\boldsymbol{x}_2\;\vert\;\dotsc\;\vert\;\boldsymbol{x}_{\frac{L}{B}}
\right]
=\text{Split}\Big(\text{flatten}(X), \frac{L}{B}\Big).
\end{equation}
Then, the block weights $W^i_B$ are multiplied with the corresponding block-wise vectors $\boldsymbol{x}^{(i)}$, respectively, to obtain the following local feature interaction vector
\begin{equation}
\begin{aligned}
\label{Eq4_3_4}
\text{reshape}\Big(H,\frac{L}{B},B\Big)=\text{reshape}\Big(\left[
\boldsymbol{x}_{1}W^1_B\;\Big\vert\;\boldsymbol{x}_{2}W^2_B\;\Big\vert\;\dotsc\;\Big\vert\;\boldsymbol{x}_{\frac{L}{B}}W^{\frac{L}{B}}_B
\right], \frac{L}{B},B\Big)=
\left[
\begin{array}{c}
\boldsymbol{x}_1W^1_B\\ 
\vdots\\ 
\boldsymbol{x}_{\frac{L}{B}}W^{\frac{L}{B}}_B
\end{array}
\right].
\end{aligned}
\end{equation}
Therefor, the output of UniMixing module can be obtained as
\begin{equation}
\label{Eq4_3_5}
\text{UniMixing}(X)=\text{reshape}\Big(W_G\text{reshape}\Big(H, \frac{L}{B}, B\Big), 1, L\Big).
\end{equation}
With this operation, compared to directly using the reconstructed matrix $W^\text{perm}$, 
the optimized computation pipeline reduces the computational cost from $O(L^2)$ to 
$O(L^2/B+LB)$, and avoids the creation of large intermediate variables during computation. 
The proof of the computation pipeline optimization of (\ref{Eq4_3_2}) is provided in Appendix \ref{ComputationPipelineOptimization}. 
According to (\ref{Eq4_3_4}) and (\ref{Eq4_3_5}), 
$W_B^i$ controls the intra-block interaction pattern, while $W_G$ controls the inter-block interaction pattern.
For the embedding inputs with dimension $L$, it is no longer required that $T=H$. Compared with the TokenMixer operation, the UniMixing module possesses more diverse local and global feature mixing patterns and interaction scales, while also retaining the advantage of being learnable and optimizable.
To fulfill the doubly stochasticity of learned permutation matrices, Sinkhorn-Knopp iteration is used to makes all elements of $W_G$ and $W_B^i$ to be positive via an exponent operator and then conducts iterative normalization that alternately rescales rows and columns to sum to 1. Besides, a temperature coefficient is introduced to control the sparsity of the parameter matrix. Finally, we employ $(W_G+W_G^\mathsf{T})/2$ and $(W_B^i+W_B^{i\mathsf{T}})/2$ to achieve the symmetry constraints of parameter matrices. The final constrained weights can be obtained by
\begin{equation}
\begin{aligned}
\label{Eq4_3_6}
&\tilde{W}_G=\frac{W_G+W_G^\mathsf{T}}{2},\;\tilde{W}_B^i=\frac{W_B^i+W_B^{i\mathsf{T}}}{2},\\
&\bar{W}_G=\text{Sinkhorn-Knopp}\Big(\frac{\tilde{W}_G}{\tau}\Big), \bar{W}^i_B=\text{Sinkhorn-Knopp}\Big(\frac{\tilde{W}_B^i}{\tau}\Big),
\end{aligned}
\end{equation}
where $\tau$ is the temperature coefficient.

Then, the residual connection and normalization module are used to process the output of the UniMixing block
\begin{equation}
    O = \text{RMSNorm}(X+\text{UniMixing}(X))
\end{equation}

\paragraph{A Unified Perspective of Heterogeneous Feature Interaction.}
Observing $V_h$ in (\ref{Eq3_01}) and $\text{reshape}(H,L/B,B)$ in (\ref{Eq4_3_4}), 
we find that if the number of blocks $L//B$ is set as $T$, and $W_V^{ih}$ and $W_B^i$ have the same dimensions, 
then $\text{reshape}(H,L/B,B)=V_h$.
This implies that the local interaction projection of UniMixer is equivalent to the value 
projection of the heterogeneous attention layer under $W^{i}_V=W_B^i$.
On the other hand, the dimension and the role of $W_G$ are the same as the attention weights, 
except that $W_G$ needs to satisfy the doubly stochasticity, sparsity, and symmetry.
The feature interaction of Wukong is based on the FM component. 
Accoriding to (\ref{wukong}), the expression of $\text{FMB}(X)$ can be rewritten as $\text{FMB}(X)=\text{reshape}(\text{MLP}(\text{LN}(\text{flatten}(XI(XI)^{\mathsf{T}}Y))))$, 
where $I$ is the identity matrix with an appropriate dimension.
Let us focus on the core feature interaction module $XI(XI)^{\mathsf{T}}Y$.  
In the attention module, when $W_Q=I$, $W_K=I$, and the value matrix does not depend on the hidden state input $X$, namely, $V_h=W_V=Y$, the Attention mechanism degenerates into the FM module. Therefore, attention-based, TokenMixer-based, and Wukong-based architecture can be unified under the following single theoretical framework
\begin{equation}
\label{Eq4_3_7}
\text{UniMixing}(X)=\text{reshape}\Bigg(\underset{\text{Global Mixing Pattern}}{\underbrace{G(X,W_G)}}
\underset{\text{Local Mixing Pattern}}{\underbrace{\left[
\begin{array}{c}
\boldsymbol{x}_1W^1_B\\ 
\vdots\\ 
\boldsymbol{x}_{\frac{L}{B}}W^{\frac{L}{B}}_B
\end{array}
\right]}}, 1, L\Bigg),
\end{equation}
where $G(X, W_G)$ is a heterogeneous feature interaction projection and measures token-to-token/block-to-block interaction strength. 
To facilitate the analysis of the differences and connections among various methods, 
we consider the single-head attention setting. Under the unified theoretical framework (\ref{Eq4_3_7}), their differences are summarized in Table \ref{Tab_0301}. For the self-attention, heterogeneous attention, and FM, the global mixing pattern $G(X, W_G)$ is obtained by computing the inner-product similarity between two tokens. The global mixing pattern of TokenMixer is independent of the input token embedding.
\begin{table}
\caption{The differences of attention-based, TokenMixer-based, FM-based methods under the unified theoretical framework.}
\label{Tab_0301}
\centering
\begin{tabular}{l|cc}
\toprule
Heterogeneous Feature Mixing &Local Mixing Pattern &Global Mixing Pattern $G(X,W_G)$\\
\midrule
Self-Attention &$XW_V$ & $\text{softmax}(\frac{(XW_Q)(XW_K)^\mathsf{T}}{\sqrt{d}})$ \\
Heterogeneous Attention &$X\tilde{W}_V$ &$\text{softmax}(\frac{(X\tilde{W}_Q)(X\tilde{W}_K)^\mathsf{T}}{\sqrt{d}})$\\
TokenMixer &$X$ &$G$\\
FM &$Y$ & $XI(XI)^\mathsf{T}$\\
\bottomrule
\end{tabular}
\end{table}

\paragraph{UniMixing-Lite.} 
As shown in Fig. \ref{perm_weight}, it can be observed that as the block granularity becomes finer, the number of local interaction parameter matrices $W_B^i$ increases, and the global interaction parameter matrix $W_G$ becomes larger. This leads to redundant local interaction patterns. Meanwhile, the larger global interaction matrix is not efficient in reducing the number of parameters.
Therefore, based on the UniMixing block, we design a lightweight UniMixing module, UniMixing-Lite, to further reduce the number of module parameters and computational cost, thereby improving the scaling efficiency of the model.

To address the problem of redundancy in the local interaction pattern, 
a basis-composed module is introduced to dynamically generates the
block-specific local mixing weight. Define a set of basis matrices for $W_B^i$ as $\{Z_\ell\}_{\ell=1}^b$ and block-specific weight vectors over these bases as $\{\boldsymbol{\omega}^i\}_{i=1}^{L//B}$, where $b$ is the number of the basis local mixing weight and $\boldsymbol{\omega}^i=[{\omega}^i_1,\dotsc,{\omega}^i_b]$. In addition, for the global interaction parameter $W_G$, we use low-rank approximation to further improve the efficiency. Then, the UniMixing-Lite module can be expressed as 
\begin{equation}
\begin{aligned}
\label{Eq4_3_8}
\text{UniMixing-Lite}(X)=\;&\text{reshape}\Big(W_{r}\text{reshape}\Big(\left[
\boldsymbol{x}_{1}W^{*1}_B\;\Big\vert\;\dotsc\;\Big\vert\;\boldsymbol{x}_{\frac{L}{B}}W^{*\frac{L}{B}}_B
\right], \frac{L}{B}, B\Big), 1, L\Big),\\
O=\;&\text{RMSNorm}(X+\text{UniMixing-Lite}(X)),
\end{aligned}
\end{equation}
where $W_r=\text{Sinkhorn-Knopp}(A_GB_G)$,  $W_B^{*i}=\text{Sinkhorn-Knopp}(\sum_{\ell=1}^b\omega_\ell^iZ_\ell)$, $A_G\in\mathbb{R}^{(L//B)\times{r}}$ and $B_G\in\mathbb{R}^{{r}\times(L//B)}$. $r$ is the rank of the low-rank approximation for $W_G$.
In the UniMixing-Lite module, we retain both the low-parameterized global interaction pattern of the TokenMixer and the local interaction capability of attention for heterogeneous features. It can simultaneously leverage the advantages of both attention-based and token-mixer-based methods.

\paragraph{Pertoken SwiGLU.} After the UniMixing block, similar to \cite{jiang2026tokenmixer}, the pertoken SwiGLU is introduced to model the feature heterogeneity among different tokens. For each input token $\boldsymbol{x}_i$, the SwiGLU formulation is given as follows
\begin{equation}
    \text{pSwiGLU}(\boldsymbol{o}_i)=W_\text{down}^i((W_\text{up}^i\boldsymbol{o}_i+b_\text{up}^i)\odot\text{Swish}(W_\text{gate}^i\boldsymbol{o}_i+b_\text{gate}^i))+b_\text{down}^i,
\end{equation}
where $W_\text{up}^i, W_\text{gate}^i\in\mathbb{R}^{B\times{nB}}$, $W_\text{down}^i\in\mathbb{R}^{nB\times{B}}$,
$b_\text{up}^i, b_\text{gate}^i\in\mathbb{R}^{nB}$,
$b_\text{down}^i\in\mathbb{R}^{B}$, $\boldsymbol{o}_i$ is the UniMixing output of the $i$-th token, and $n$ is a 
hyperparameter.

\subsection{SiameseNorm}
The current RankMixer architecture \cite{zhu2025rankmixer} lacks specialized design for deep architectures, which is generally reflected in the limited effectiveness of scaling along the model depth. Although TokenMixer-Large \cite{jiang2026tokenmixer} attempts to address this problem by incorporating interval residuals and the auxiliary loss within the TokenMixer-Large Block, it does not address the root of the problem. To achieve the training stability and performance gains as model depth increases, SiameseNorm \cite{li2026siamesenormbreakingbarrierreconciling} is introduced into the UniMixer architecture as shown in Fig. \ref{framework}. As mentioned in \cite{li2026siamesenormbreakingbarrierreconciling}, SiameseNorm resolves the tension between Pre-Norm and Post-Norm by introducing two coupled streams per layer. In this subsection, these two coupled streams are denoted as $\bar{X}_i$ and $\bar{Y}_i$, which is initialized by the input embeddings $\bar{X}_0=\bar{Y}_0=X$. For the $\ell$-th block, SiameseNorm conduct the following update:
\begin{equation}
    \begin{aligned}
        \tilde{Y}_\ell=\;&\text{RMSNorm}(\bar{Y}_\ell),\quad{O}_\ell=\text{UniMixer}(\bar{X}_\ell+\tilde{Y}_\ell)\\
        \bar{X}_{\ell+1}=\;&\text{RMSNorm}(\bar{X}_\ell+O_\ell),\quad\bar{Y}_{\ell+1}=\bar{Y}_\ell+O_\ell.\nonumber
    \end{aligned}
\end{equation}
For the $M$-th UniMixer block, $\bar{X}_\ell$ and $\bar{Y}_\ell$ are fused to generate the final representation, which is formulated as
\begin{equation}
    X_\text{output}=\bar{X}_M+\text{RMSNorm}(\bar{Y}_M).
\end{equation}


\subsection{UniMixer Training Strategies}
To require sparsity in the parameter matrices $W_G$ and $W_B^i$, we introduce a temperature coefficient to control their sparsity level. However, a smaller temperature leads to sparser weights, while also resulting in the gradients to become sparse, weak, or even unstable. This can make the training process difficult, and optimization get trapped in local optima. On the other hand, our experiments show that the sparsity of the weight parameters has a significantly positive effect on model performance, as shown in Table \ref{tab:ablation} (subsection 5.3). Therefore, such sparsity is indispensable. A commonly used approach is to apply linear temperature annealing during the training process: starting with a relatively high initial temperature (e.g., $\tau = 1.0$) and gradually annealing it linearly to 0.05 as the number of training iterations increases, which is formulated as
\begin{equation}
    \tau_j=\max\Big\{\tau_\text{start}-\frac{(\tau_\text{start}-\tau_\text{end})j}{J}, \tau_\text{end}\Big\},
\end{equation}
where $\tau_j$ is the $j$-th temperature coefficient, $\tau_\text{start}$ and $\tau_\text{end}$ are the initial temperature and final temperature, respectively, $J$ is the iteration range for temperature annealing. When the amount of data is insufficient, linear annealing may lead to inadequate exploration in the early stage with a high temperature coefficient, or suboptimal optimization in the later stage with a low temperature coefficient. To address this, we can first use a high temperature coefficient (e.g., $\tau = 1.0$) to cold-start the training of the model; once the model is well trained, we then lower the temperature coefficient (e.g., $\tau = 0.05$) and retrain the low-temperature model using the weights of the high-temperature model as initialization.

%% file: sections/Experiments.tex
\section{Experiments}
\label{Experiments}
In this section, we conduct extensive experiments to compare the performance of the present UniMixer architecture with existing state-of-the-art (SOTA) approaches and to answer the following questions: 
\begin{itemize}
    \item [\textbf{Q1}:] Does the scaling efficiency of the UniMixer architecture outperform the SOTA architecture?
    \item [\textbf{Q2}:] How does the performance of the proposed method change under different settings of global and local mixing pattern?
    \item [\textbf{Q3}:] Does the lightweigh module, UniMixing-Lite, further improve the scaling efficiency?
    \item [\textbf{Q4}:] When deployed in a real-world online system, does UniMixer/UniMixing-Lite improve business metrics in A/B testing?
\end{itemize}

\subsection{Experimental Setup}
\paragraph{Datasets and Evaluation Metrics.}
We use the logged data from the real-world training dataset of the advertising delivery scenario on Kuaishou to model user retention and conduct the offline and online evaluation. The dataset contains over 0.7 billion user samples collected over one year, which comprises hundreds of heterogeneous features such as numerical features, ID features, cross features, and sequential features. A binary label (User Retention = 1/0) indicates whether the user returns to the Kuaishou application on the day following the users' first activation.
For the scaling evaluation metrics of recommendation models, we adopt the two common metrics used in recommender system, i.e., area under the ROC curve (AUC), and UAUC (User-Level AUC), to evaluate the model performance, and dense parameter count, FLOPs, and MFU to evaluate the model efficiency.

\paragraph{Baselines and Experimental Details.} We compare the present 2-blocks/4-blocks UniMixer/UniMixing-Lite architectures with the following representative SOTA frameworks, categorized by modeling paradigm
\begin{itemize}
    \item \textbf{Attention-Based Architectures}: Heterogeneous Attention \cite{gui2023hiformer}, HiFormer \cite{gui2023hiformer}, and FAT \cite{yan2025scaling}, which use the field-specific query, key, and value projections to achieve heterogeneous feature interaction.
    \item \textbf{TokenMixer-Based Framework}: RankMixer \cite{zhu2025rankmixer} and TokenMixer-Large \cite{jiang2026tokenmixer}, which employ the rule-based token mixing operation to perform the feature interaction.
    \item \textbf{FM-Based Framework}: Wukong \cite{zhang2024wukong}, which concatenates the outputs of a FMB and a linear projection layer to upscale the interaction component.
\end{itemize}
All experiments are conducted in a hybrid distributed training framework composed of 40 GPUs. All models use consistent optimizer hyperparameters: both the dense and sparse parts are optimized with Adam, with a learning rate set to 0.001.

\subsection{Performance Comparison (for Q1)}
The SOTA scaling architectures with approximately 100 million parameters are used to compare with UniMixer and UniMixer-Lite to explore their scaling laws. The heterogeneous attention architecture is used to be the base model. The main performance results of our models and the SOTA models are provided in Table \ref{tab:performance}. It can be observed that, under smaller parameter budgets and computational costs, both UniMixer and UniMixer-Lite architectures significantly outperform other SOTA models across multiple metrics. 
\begin{table}[ht]
\centering
\caption{Performance and efficiency of $\sim100$M-parameter \textbf{UniMixer} and SOTA models in ad serving scenarios.}
\label{tab:performance}
\resizebox{\linewidth}{!}{
\begin{tabular}{lcccccc}
\toprule
\multirow{2}{*}{\textbf{Model}} 
& \multicolumn{4}{c}{\textbf{User Retention}} 
& \multicolumn{2}{c}{\textbf{Efficiency}} \\
\cmidrule(lr){2-5} \cmidrule(lr){6-7}
& AUC$\uparrow$ & $\Delta$AUC$\uparrow$ & UAUC$\uparrow$ & $\Delta$UAUC$\uparrow$ & {Params} & {FLOPs/Batch} \\
\midrule
Heterogeneous Attention \cite{gui2023hiformer} & 0.744577 & -- & 0.733829 & -- & 132.7M & 1.68T \\
\midrule
HiFormer \cite{gui2023hiformer} & 0.741685 & -0.2892\% & 0.731086 & -0.2743\% & 107.5M & 1.37T \\
Wukong \cite{zhang2024wukong} & 0.744477 & -0.0100\% & 0.733849 & 0.0020\% & 107.1M & 1.40T \\
FAT \cite{yan2025scaling} &0.744883 & 0.0306\% &0.734280 & 0.0451\%& 138.4M & 1.83T \\
RankMixer \cite{zhu2025rankmixer} &0.749329 & 0.4752\%&0.738938 & 0.5109\% & 135.5M & 1.68T \\
TokenMixer-Large \cite{jiang2026tokenmixer} & 0.748410 & 0.3833\%& 0.737940 & 0.4111\% & 103.3M & 1.27T \\
\midrule
\textbf{UniMixer-2-Blocks 67.5M (Ours)} &{0.749770} & {0.5193\%} &{0.739331} & {0.5502\%} & 67.5M & 2.07T \\
\textbf{UniMixer-2-Blocks 101.5M (Ours)} &{0.750238}  & {0.5661\%} &{0.739983} &{0.6154\%} & 101.5M & 2.50T \\
\textbf{UniMixer-Lite-2-Blocks 42.4M (Ours)} &\underline{0.751121}  & \underline{0.6544\%} &\underline{0.740739} &\underline{0.6910\%}& 42.4M & 2.17T \\
\textbf{UniMixer-Lite-2-Blocks 76.2M (Ours)} &\underline{0.751401}  & \underline{0.6824\%} &\underline{0.741215} &\underline{0.7386\%} & 76.2M & 2.60T \\
\textbf{UniMixer-Lite-4-Blocks 38.2M (Ours)} &\textbf{0.752327}  & \textbf{0.7750\%} &\textbf{0.742091} & \textbf{0.8190\%}& 38.2M & 1.26T \\
\textbf{UniMixer-Lite-4-Blocks 84.5M (Ours)} &\textbf{0.752718}  & \textbf{0.8141\%} &\textbf{0.742530} & \textbf{0.8701\%}& 84.5M & 4.24T \\

\bottomrule
\end{tabular}}
\end{table}

\begin{figure}[ht]
\begin{center}
\centerline{\includegraphics[width=0.9\columnwidth]{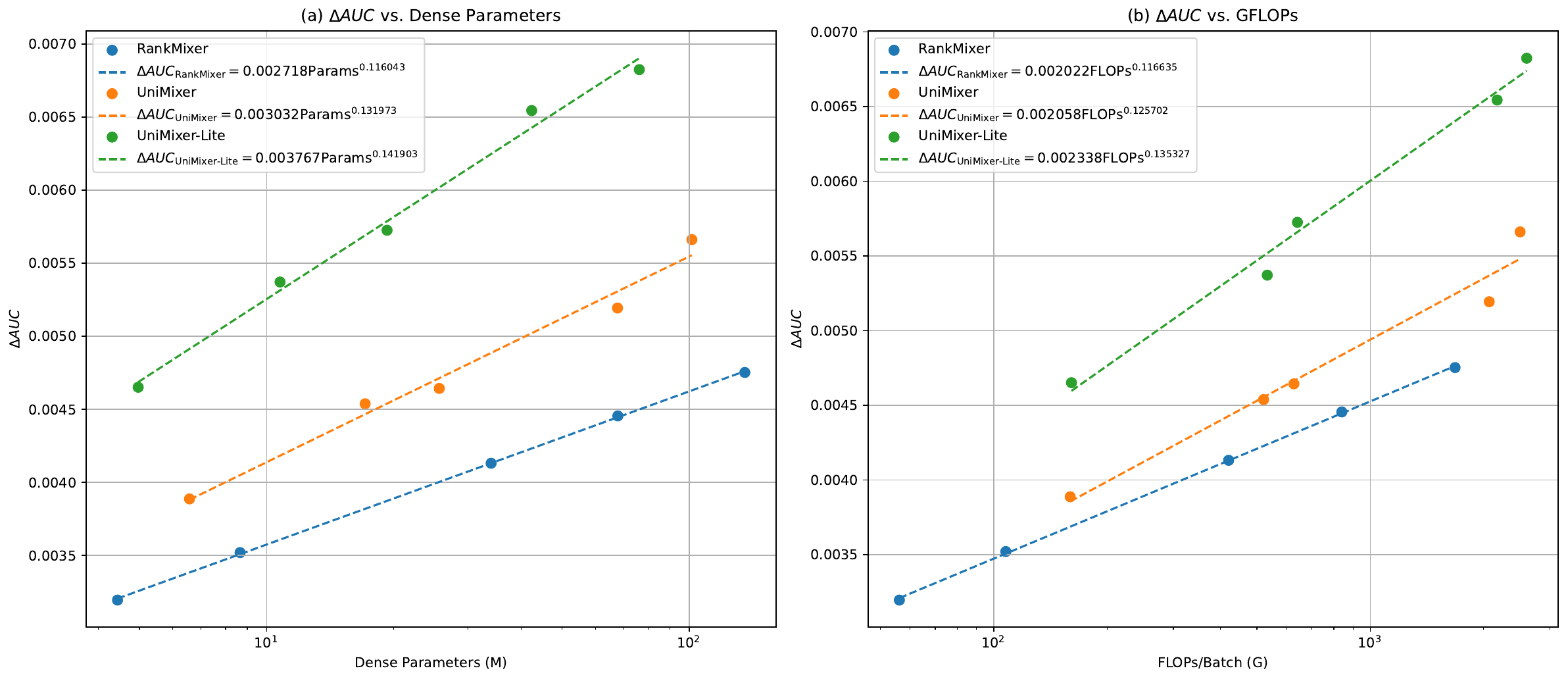}}
\caption{The scaling laws between AUC and Parameters/FLOPs for UniMixer-2-Blocks/UniMixer-Lite-2-Blocks and RankMixer architectures. The x-axis is presented on a logarithmic scale.}
\label{ScalingCurves}
\end{center}
\end{figure}

Subsequently, in this advertising delivery scenario, the performance of RankMixer outperforms all other SOTA models except UniMixer/UniMixer-Lite. Therefore, we select the strongest SOTA model together with UniMixer/UniMixer-Lite for a scaling laws comparison. All models are trained on the same dataset with consistent hyperparameters. Their scaling curves with respect to parameters and FLOPs are given in Figure \ref{ScalingCurves}. As observed, the AUC of all three models exhibits clear power-law trends as the number of parameters/FLOPs increases. UniMixer-Lite achieve the best scaling efficiency and exhibits a steeper improvement slope. According to the relationship between parameter count and AUC as shown in Fig. \ref{ScalingCurves}, the well-behaved scaling laws between AUC and Parameters/FLOPs for RankMixer, 
UniMixer and UniMixer-Lite can be formulated as follows
\begin{equation}
\begin{aligned}
\Delta\text{AUC}_\text{RankMixer}=&\;0.002718\text{Params}^{0.116043},\;\Delta\text{AUC}_\text{RankMixer}=0.002022\text{FLOPs}^{0.116635},\\
\Delta\text{AUC}_\text{UniMixer}=&\;0.003032\text{Params}^{0.131973},\;\Delta\text{AUC}_\text{UniMixer}=0.002058\text{FLOPs}^{0.125702},\\
\Delta\text{AUC}_\text{UniMixer-Lite}=&\;0.003767\text{Params}^{0.141903},\;\Delta\text{AUC}_\text{UniMixer-Lite}=0.002338\text{FLOPs}^{0.135327}.\nonumber
\end{aligned}
\end{equation}
Among two constants in the scaling laws, the scaling exponent constant has the most significant impact on performance growth, which is the dominant factor in scaling efficiency. UniMixer-Lite demonstrates the strongest scaling efficiency, achieving both the largest scaling exponent and coefficient across parameters and FLOPs. This indicates that it benefits the most from increased model capacity.

\subsection{Ablation Studies (for Q2)}
\label{AblationStudies}
To explore the properties of global and local mixing weights, as well as the contribution of each module in UniMixer to AUC gains, we conduct ablation studies on various UniMixer variants and measure their relative AUC changes compared to the full UniMixer model. All variants are trained under similar settings. The results are shown in Table \ref{tab:ablation}, which illustrate that removing any module or violating any parameter constraints leads to performance degradation, with low temperature coefficient and model warm-up having the most significant impact on overall performance.
\begin{table}[ht]
\centering
\caption{Ablation on components of UniMixer 6.57M.}
\label{tab:ablation}
\resizebox{\linewidth}{!}{
\begin{tabular}{lcccccc}
\toprule
\multirow{2}{*}{\textbf{Setting}} 
& \multicolumn{4}{c}{\textbf{User Retention}} 
& \multicolumn{2}{c}{\textbf{Efficiency}} \\
\cmidrule(lr){2-5} \cmidrule(lr){6-7}
& AUC$\uparrow$ & $\Delta$AUC$\uparrow$ & UAUC$\uparrow$ & $\Delta$UAUC$\uparrow$ & {Params} & {FLOPs/Batch} \\
\midrule
\textbf{UniMixer} &0.748464  & -- &0.738017 & -- & 6.57M & 0.16T \\
w/o Temperature Coefficient &0.746819 &-0.1645\% &0.736527 & -0.1490\% & 6.57M & 0.16T \\
w/o Symmetry Constraint &0.747891 &-0.0573\% &0.737447 &-0.0570\% & 6.57M & 0.16T\\
w/o Block-Specific Local Mixing Weight  &0.748028 & -0.0436\% &0.737770 & -0.0240\%& 6.53M & 0.16T \\
w/o Model Warm-Up &0.747608 & -0.0856\% &0.737180 & -0.0837\%& 6.57M & 0.16T \\
SiameseNorm $\to$ Post Norm &0.748191 & -0.0273\% &0.737660 & -0.0357\%& 6.56M & 0.16T \\
\bottomrule
\end{tabular}}
\end{table}

\subsection{Performance of the UniMixing-Lite Module (for Q3)}
\begin{figure}[ht]
\begin{center}
\centerline{\includegraphics[width=0.8\columnwidth]{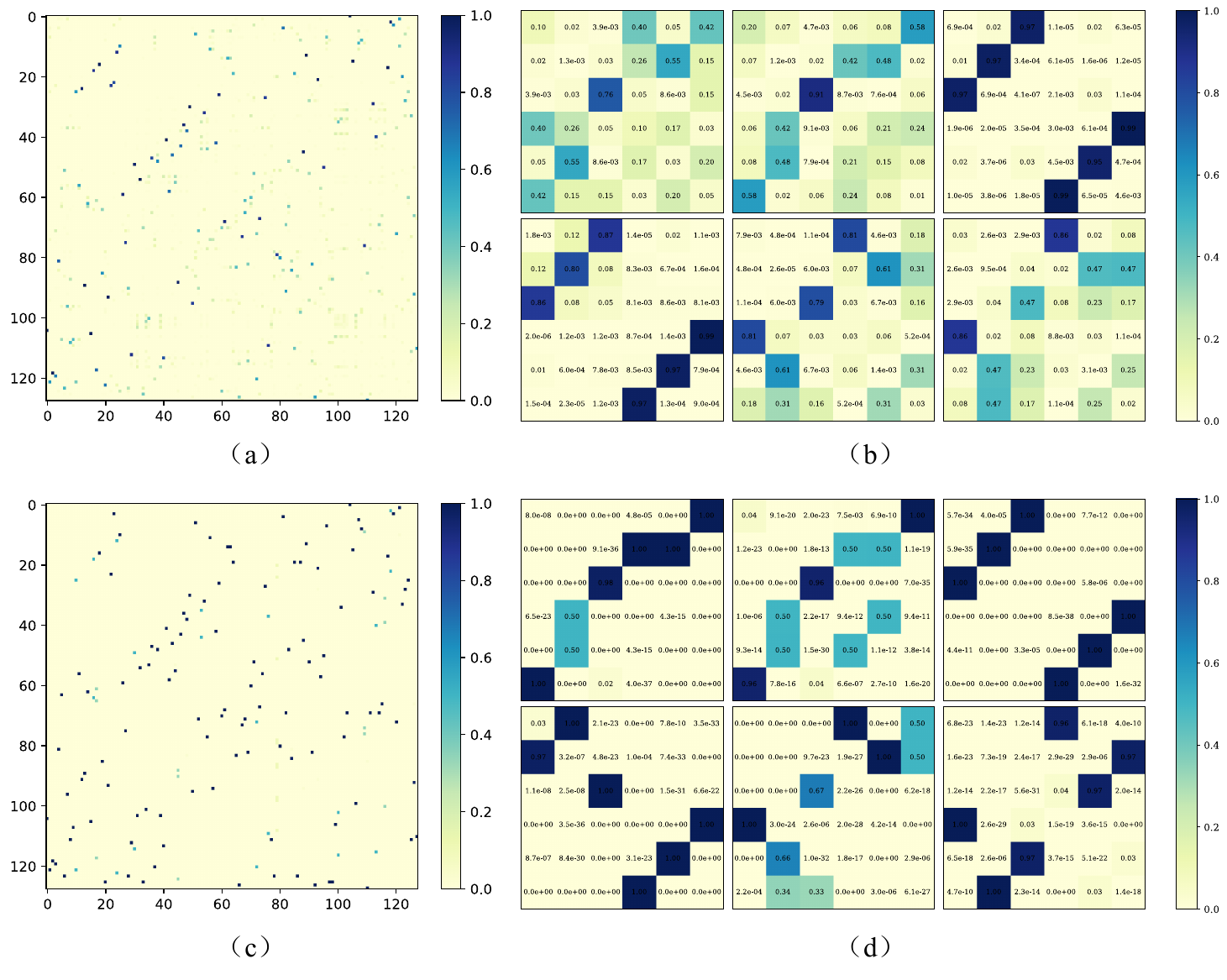}}
\caption{$\bar{W}_G$ and $\bar{W}_B^i$ of UniMixer-Lite with different temperature coefficients. (a) $\bar{W}_G$ with $\tau=1$; (b) $\{\bar{W}_B^i\}_{i=1}^6$ with $\tau=1$; (c) $\bar{W}_G$ with $\tau=0.05$; (d) $\{\bar{W}_B^i\}_{i=1}^6$ with $\tau=0.05$.}
\label{global_local_mixing_weights}
\end{center}
\end{figure}

According to the scaling trends from Fig. \ref{ScalingCurves}, it can be obaserved that the present UniMixing-Lite architecture possesses the best parameter efficiency and computational efficiency. Here, we conduct experiments to investigate the effects of different basis numbers $b$ for $\{Z_\ell\}_{\ell=1}^b$, different rank $r$ for $A_G$ and $B_G$ and different UniMixer block number. As shown in Table \ref{tab:UniMixing_Lite}, as the basis numbers $b$ and the rank $r$ for $A_G,B_G$ increase, the model performance improves accordingly. However, in terms of parameter efficiency, increasing the number of basis $b$  yields a higher AUC gain than increasing the rank $r$. 
To observe the effects of the low-rank approximation $A_GB_G$ and basis matrices $\{Z_\ell\}_{\ell=1}^b$ with the Sinkhorn–Knopp operation on reconstructing the global and local mixing matrices, in a 2-blocks-UniMixer-Lite architecture, we visualize the reconstructed global matrix $\bar{W}_G$ and the first six local mixing matrices $\bar{W}^i_B$ of the first UniMixer block with the temperature coefficients $\tau=1$ and $\tau=0.05$, as shown in Fig. \ref{global_local_mixing_weights}. The input embedding dimension is 768, and the block size is 6; therefore, we have $\bar{W}_G\in\mathbb{R}^{128\times128}$ and $\bar{W}^i_B\in\mathbb{R}^{6\times6}$, where $A_G\in\mathbb{R}^{128\times16}$ and $B_G\in\mathbb{R}^{16\times128}$. According to Fig. \ref{global_local_mixing_weights}, although the low-rank approximation and basis matrices are used in the module, the Sinkhorn–Knopp operation can still ensure that the matrix remains close to full rank. In addition, compared with Figs. \ref{global_local_mixing_weights}(a)(b) and (c)(d), the global and local mixing matrices with a lower temperature coefficient exhibit sharper interaction distributions than those with a higher temperature coefficient. From the ablation results given in the ablation studies, we can conclude that the sparsity of $\bar{W}_G$ and $\bar{W}^i_B$ leads to a significant improvement in model performance.

\begin{figure}[ht]
\begin{center}
\centerline{\includegraphics[width=0.9\columnwidth]{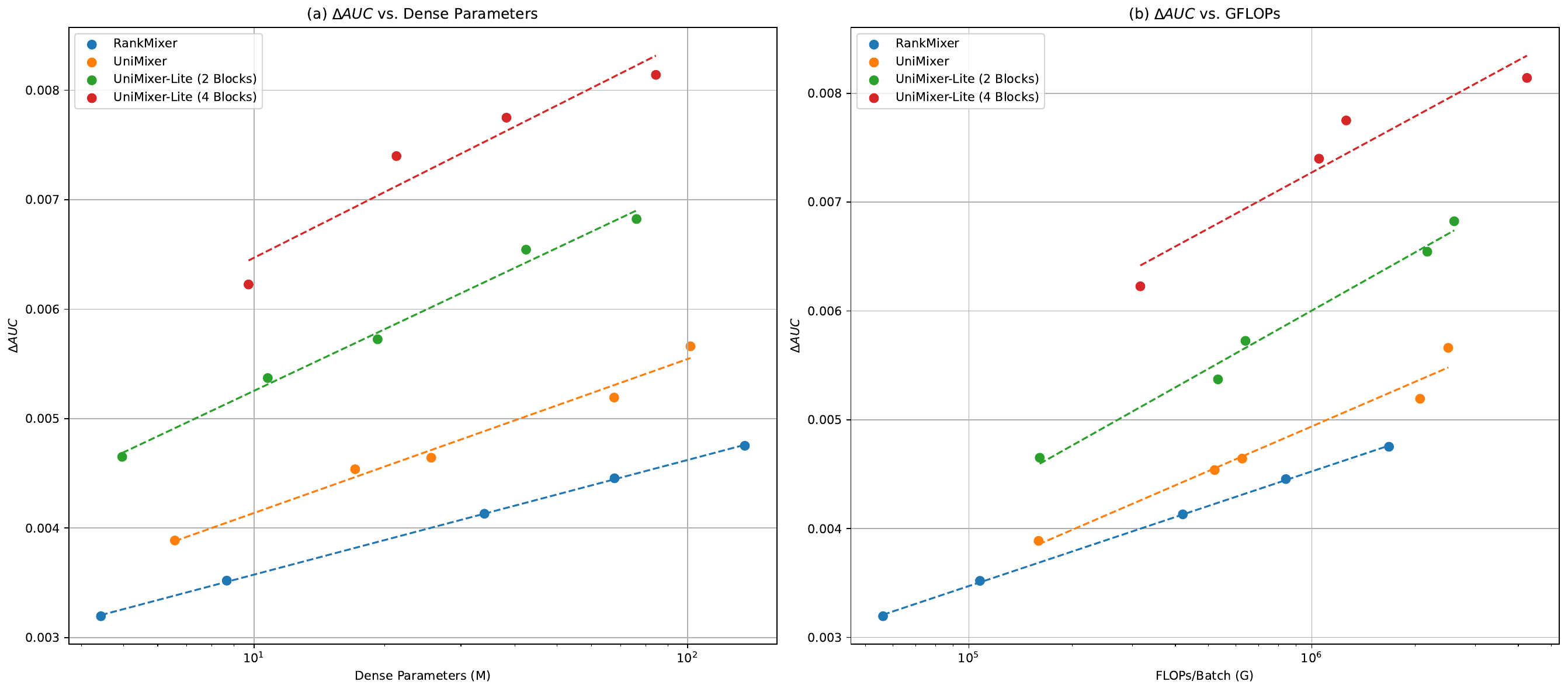}}
\caption{The scaling curves between AUC and Parameters/FLOPs for RankMixer and UniMixer/UniMixer-Lite with 2 blocks and 4 blocks.}
\label{ScalingCurves_4blocks}
\end{center}
\end{figure}

On the other hand, according to Table \ref{tab:UniMixing_Lite}, it can be observed that as the depth of UniMixer increases, the developed model continues to exhibit a clear scaling-up trend, whereas RankMixer shows a performance degradation as the RankMixer blocks are stacked. The scaling curves of UniMixing-Lite with 2 blocks and 4 blocks are shown in Fig. \ref{ScalingCurves_4blocks}, which implies that scaling along depth is more efficient than scaling along width.
\begin{table}[ht]
\centering
\caption{Effects of the basis number, rank, and UniMixer block number in UniMixing-Lite.}
\label{tab:UniMixing_Lite}
\resizebox{\linewidth}{!}{
\begin{tabular}{lcccccc}
\toprule
\multirow{2}{*}{\textbf{Basis Number}} 
& \multicolumn{4}{c}{\textbf{User Retention}} 
& \multicolumn{2}{c}{\textbf{Efficiency}} \\
\cmidrule(lr){2-5} \cmidrule(lr){6-7}
& AUC$\uparrow$ & $\Delta$AUC$\uparrow$ & UAUC$\uparrow$ & $\Delta$UAUC$\uparrow$ & {Params} & {FLOPs/Batch} \\
\midrule
b=2 &0.749228 & -- & 0.738776 & -- & 4.968M & 0.161T \\
b=4 &0.750230 & 0.1002\% &0.739876 & 0.1100\% & 4.973M & 0.161T \\
b=8 &0.750283 & 0.1055\% &0.739854 & 0.1078\% & 4.98M & 0.161T \\
\midrule
\midrule
\multirow{2}{*}{\textbf{Rank}} 
& \multicolumn{4}{c}{\textbf{User Retention}} 
& \multicolumn{2}{c}{\textbf{Efficiency}} \\
\cmidrule(lr){2-5} \cmidrule(lr){6-7}
& AUC$\uparrow$ & $\Delta$AUC$\uparrow$ & UAUC$\uparrow$ & $\Delta$UAUC$\uparrow$ & {Params} & {FLOPs/Batch} \\
\midrule
r=2 &0.748568 & -- &0.738177 & -- & 4.4525M & 0.160T \\
r=64 &0.749002 & 0.0434\% &0.738604 & 0.0427\% & 4.705M & 0.160T \\
r=128 &0.749228 & 0.0660\% & 0.738776 & 0.0599\% & 4.9675M & 0.161T \\
r=256 &0.749539 & 0.0971\% &0.739437 & 0.1260\% & 5.4925M & 0.163T \\
\midrule
\midrule
\multirow{2}{*}{\textbf{Block Number}} 
& \multicolumn{4}{c}{\textbf{User Retention}} 
& \multicolumn{2}{c}{\textbf{Efficiency}} \\
\cmidrule(lr){2-5} \cmidrule(lr){6-7}
& AUC$\uparrow$ & $\Delta$AUC$\uparrow$ & UAUC$\uparrow$ & $\Delta$UAUC$\uparrow$ & {Params} & {FLOPs/Batch} \\
\midrule
RankMixer-2-Blocks   &0.747772 & -- &0.737322  & -- & 4.4375M & 0.056T \\
RankMixer-4-Blocks   &0.746706 & -0.1066\% &0.736018  & -0.1304\% & 8.6575M & 0.108T \\
\midrule
UniMixer-Lite-2-Blocks   &0.749228 & -- & 0.738776 & -- & 4.9675M & 0.161T \\
UniMixer-Lite-4-Blocks &0.750803 & 0.1575\% &0.740594 & 0.1818\% & 9.715M & 0.316T \\
UniMixer-Lite-8-Blocks &0.750875 & 0.1647\% &0.740602 & 0.1826\% & 19.2125M & 0.629T \\
\bottomrule
\end{tabular}}
\end{table}

\subsection{Online A/B Test Results (for Q4)}
To verify the online performance of the proposed UniMixer architecture, we have deployed UniMixer and UniMixer-Lite across multiple advertising delivery scenarios on Kuaishou. In the online A/B test, we measure user engagement using the Cumulative Active Days (CAD) over a 30-day observation window, excluding the installation day (day 0). Across multiple scenarios, CAD of D1-D30 increased by more than $15\%$ on average.

%% file: sections/Conclusions.tex
\section{Conclusions}
\label{Conclusions}
In this work, a unified scaling framework is established for scaling laws in recommendation systems, which ridges the connections among attention-based, TokenMixer-based, and FM-based methods and makes it possible to leverage their respective strengths. From the obtained scaling laws, compared with the SOTA architectures, the present UniMixer-Lite achieved the best parameter efficiency and computational efficiency. We have deployed the architectures across multiple scenarios at Kuaishou, yielding significant offline and online gains. This work no longer treats existing scaling blocks (e.g., Heterogeneous Attention, TokenMixer, Wukong) in recommender in isolation. Instead, it establishes a unified theoretical framework that provides guidance for scaling design in recommendation systems. We believe that the unified architecture can help the recommendation systems community achieve its own ``attention moment''. The unified module, UniMixer, serves as a fundamental block tailored for the recommendation domain, whose applicability can be further extended to user behavior sequence modeling and generative recommendation tasks.

%% file: sections/Appendix.tex
\section{A numerical example of equivalent transformation of TokenMixer}
\label{NumericalExample}
The following input hidden state $X\in\mathbb{R}^{2\times{6}}$ is given
\begin{equation}
X=\left[
\begin{array}{ccc:ccc}
x_1 & x_2 & x_3 & x_4 & x_5 & x_6\\ 
x_7 & x_8 & x_9 & x_{10} & x_{11} & x_{12}
\end{array}
\right],\nonumber
\end{equation}
where $x_i$ is a scalar.
Then the input hidden state $X$ passed through the TokenMixer operation is transformed into
\begin{equation}
\text{TokenMixer}(X)=\left[
\begin{array}{cccccc}
x_1 & x_2 & x_3 & x_7 & x_8 & x_9\\ 
\hdashline
x_4 & x_5 & x_6 & x_{10} & x_{11} & x_{12}
\end{array}
\right].\nonumber
\end{equation}
The output of TokenMixer can be flatten as a vector 
\begin{equation}
    \label{appendix_01}
    \text{flatten}(\text{TokenMixer}(X))=[x_1, x_2, x_3, x_7, x_8, x_9, x_4, x_5, x_6, x_{10}, x_{11}, x_{12}]^{\mathsf{T}}
\end{equation}
On the other hand, the vector $\text{flatten}(X)$ can be transformed into $\text{flatten}(\text{TokenMixer}(X))$ by multiplying a $12\times12$ matrix, which can be formulated as 
\begin{equation}
\label{appendix_02}
\underset{\text{$W^\text{perm}$}}{\underbrace{
\left[
\begin{array}{ccc:ccc:ccc:ccc}
1 & 0 &0  & 0 &0 & 0 &0  & 0 &0 & 0 &0  & 0\\
 0 & 1 &0  & 0 &0 & 0 &0  & 0 &0 & 0 &0  & 0\\
 0 & 0 &1  & 0 &0 & 0 &0  & 0 &0 & 0 &0  & 0\\
 \hdashline
 0 & 0 &0  & 0 &0 & 0 &1  & 0 &0 & 0 &0  & 0 \\
 0 & 0 &0  & 0 &0 & 0 &0  & 1 &0 & 0 &0  & 0 \\
  0 & 0 &0  & 0 &0 & 0 &0  & 0 &1 & 0 &0  & 0 \\
\hdashline
 0 & 0 &0  & 1 &0 & 0 &0  & 0 &0 & 0 &0  & 0 \\
 0 & 0 &0  & 0 &1 & 0 &0  & 0 &0 & 0 &0  & 0 \\
 0 & 0 &0  & 0 &0 & 1 &0  & 0 &0 & 0 &0  & 0 \\
 \hdashline
0 & 0 &0  & 0 &0 & 0 &0  & 0 &0 & 1 &0  & 0 \\
0 & 0 &0  & 0 &0 & 0 &0  & 0 &0 & 0 &1  & 0 \\
0 & 0 &0  & 0 &0 & 0 &0  & 0 &0 & 0 &0  & 1 \\
\end{array}
\right]}}\quad
\underset{\text{fllaten}(X)}{\underbrace{
\left[
\begin{array}{c}
x_1\\ 
x_2\\ 
x_3\\ 
x_4\\
x_5\\
x_6\\ 
x_7\\
x_8\\
x_9\\ 
x_{10}\\ 
x_{11}\\ 
x_{12}
\end{array}
\right]}}
=
\underset{\text{flatten}(\text{TokenMixer}(X))}{\underbrace{
\left[
\begin{array}{c}
x_1\\ 
x_2\\ 
x_3\\ 
x_7\\
x_8\\
x_9\\ 
x_4\\
x_5\\
x_6\\ 
x_{10}\\ 
x_{11}\\ 
x_{12}
\end{array}
\right]}}
\end{equation}
According to (\ref{appendix_01}) and (\ref{appendix_02}), the TokenMixer operation in this numerical example is equivalently transformed into the form of multiply matrix. 
In addition, the permutation Matrix $W^\text{perm}\in\mathbb{R}^{12\times12}$ can be equivalently decomposed into the Kronecker product of the follwing two small matrices
\begin{equation}
W^\text{perm}=
\underset{\text{Global Mixing Matrix}}{\underbrace{
\left[
\begin{array}{cccc}
1 & 0 & 0 & 0 \\ 
0 & 0 & 1 & 0 \\ 
0 & 1 & 0 & 0 \\ 
0 & 0 & 0 & 1
\end{array}
\right]}}
\otimes
\underset{\text{Local Mixing matric}}{\underbrace{
\left[
\begin{array}{cccc}
1 & 0 & 0  \\ 
0 & 1 & 0  \\ 
0 & 0 & 1
\end{array}
\right]}}
.\nonumber
\end{equation}

\section{The computation pipeline optimization of the UniMixing module}
\label{ComputationPipelineOptimization}
Define $W_G\in\mathbb{R}^{(L//B)\times{L//B}}$ and $W_B^i\in\mathbb{R}^{}$ as follows
\begin{equation}
W_G=\left[
\begin{array}{ccc}
w^G_{(1,1)} & \dotsc &w^G_{(1,L//B)}\\ 
\dotsc & \dotsc & \dotsc\\ 
w^G_{(L//B,1)} & \dotsc & w^G_{(L//B,L//B)}
\end{array}
\right],
W_B^{i}=
\left[
\begin{array}{ccc}
v^{i}_{(1,1)} & \dotsc &v^{i}_{(1,B)}\\ 
\dotsc & \dotsc & \dotsc\\ 
v^{i}_{(B,1)} & \dotsc & v^{i}_{(B,B)}
\end{array}
\right],
\end{equation}
where $w_{ij}$ and $v_{ij}$ are the scalars. According to (\ref{Eq4_3_3}), 
$\text{flatten}(X)$ is evenly split into $L//B$ vectors, which is rewritten as
\begin{equation}
\text{flatten}(X)=
\left[
\boldsymbol{x}_1\vert\;\boldsymbol{x}_2\;\vert\;\dotsc\;\vert\;\boldsymbol{x}_{\frac{L}{B}}
\right]^\mathsf{T},
\end{equation}
where $\boldsymbol{x}_i$ is a row vector of dimension $B$.

According to the origin expression of UniMing (\ref{Eq4_3_2}), the term $\Big(W_G\otimes\{W^i_B\}_{i=1}^{L//B}\Big)\text{flatten}(X)$ can be rewritten as 
\begin{equation}
\begin{aligned}
\label{appendix03}
\Big(W_G\otimes\{W^i_B\}_{i=1}^{L//B}\Big)\text{flatten}(X)
=&\left[
\begin{array}{ccc}
w^G_{(1,1)}W_B^{1} & \dotsc &w^G_{(1,L//B)}W_B^{L//B}\\ 
\dotsc & \dotsc & \dotsc\\ 
w^G_{(L//B,1)}W_B^{1} & \dotsc & w^G_{(L//B,L//B)}W_B^{L//B}
\end{array}
\right]
\left[
\begin{array}{c}
\boldsymbol{x}^\mathsf{T}_1\\ 
\vdots\\ 
\boldsymbol{x}^\mathsf{T}_{\frac{L}{B}}
\end{array}
\right]\\
=&
\left[
\begin{array}{c}
w^G_{(1,1)}W_B^{1}\boldsymbol{x}^\mathsf{T}_1 + \dotsc + w^G_{(1,L//B)}W_B^{L//B}\boldsymbol{x}^\mathsf{T}_{\frac{L}{B}}\\ 
\dotsc\\ 
w^G_{(L//B,1)}W_B^{1}\boldsymbol{x}^\mathsf{T}_1 + \dotsc + w^G_{(L//B,L//B)}W_B^{L//B}\boldsymbol{x}^\mathsf{T}_{\frac{L}{B}} 
\end{array}
\right]\in\mathbb{R}^{L\times1}
\end{aligned}
\end{equation}

On the other hand, we can obtain the following expression
\begin{equation}
\label{appendix04}
\begin{aligned}
W_G\text{reshape}&\Big(\left[
\boldsymbol{x}_{1}W^{1\mathsf{T}}_B\;\Big\vert\;\dotsc\;\Big\vert\;\boldsymbol{x}_{\frac{L}{B}}W^{\frac{L}{B}\mathsf{T}}_B
\right], \frac{L}{B},B\Big)\\
=&\left[
\begin{array}{ccc}
w^G_{(1,1)} & \dotsc &w^G_{(1,L//B)}\\ 
\dotsc & \dotsc & \dotsc\\ 
w^G_{(L//B,1)} & \dotsc & w^G_{(L//B,L//B)}
\end{array}
\right]
\left[
\begin{array}{c}
\boldsymbol{x}_1W^{1\mathsf{T}}_B\\ 
\vdots\\ 
\boldsymbol{x}_{\frac{L}{B}}W^{\frac{L}{B}\mathsf{T}}_B
\end{array}
\right]\\
=&
\left[
\begin{array}{c}
w^G_{(1,1)}\boldsymbol{x}_1W^{1\mathsf{T}}_B+\dotsc+w^G_{(1,L//B)}\boldsymbol{x}_{\frac{L}{B}}W^{\frac{L}{B}\mathsf{T}}_B\\ 
\dotsc\\ 
w^G_{(L//B,1)}\boldsymbol{x}_1W^{1\mathsf{T}}_B+\dotsc+w^G_{(L//B,L//B)}\boldsymbol{x}_{\frac{L}{B}}W^{\frac{L}{B}\mathsf{T}}_B
\end{array}
\right]\in\mathbb{R}^{\frac{L}{B}\times{B}}
\end{aligned}
\end{equation}
The element in ( (\ref{appendix03})) and the element in  (\ref{appendix04}) satisfy
\begin{equation}
w^G_{(i,1)}W_B^{1}\boldsymbol{x}^\mathsf{T}_1 + \dotsc + w^G_{(i,L//B)}W_B^{L//B}\boldsymbol{x}^\mathsf{T}_{\frac{L}{B}}=
(w^G_{(i,1)}\boldsymbol{x}_1W^{1\mathsf{T}}_B+\dotsc+w^G_{(i,L//B)}\boldsymbol{x}_{\frac{L}{B}}W^{\frac{L}{B}\mathsf{T}}_B)^{\mathsf{T}},
\end{equation}
which results in
\begin{equation}
\Big(W_G\otimes\{W^i_B\}_{i=1}^{L//B}\Big)\text{flatten}(X)=
\text{reshape}\Big(W_G\text{reshape}\Big(\left[
\boldsymbol{x}_{1}W^{1\mathsf{T}}_B\;\Big\vert\;\dotsc\;\Big\vert\;\boldsymbol{x}_{\frac{L}{B}}W^{\frac{L}{B}\mathsf{T}}_B
\right], \frac{L}{B},B\Big),L,1\Big)\nonumber
\end{equation}
Since both $W_B^{i}$ and $W_B^{i\mathsf{T}}$ are learnable parameters, the transpose of the parameter does not affect the model. Therefore, the UniMixing module after the computation pipeline optimization can be formulated as
\begin{equation}
 \text{UniMixing}(X)=\text{reshape}\Big(W_G\text{reshape}\Big(\left[
\boldsymbol{x}_{1}W^1_B\;\Big\vert\;\boldsymbol{x}_{2}W^2_B\;\Big\vert\;\dotsc\;\Big\vert\;\boldsymbol{x}_{\frac{L}{B}}W^{\frac{L}{B}}_B
\right], \frac{L}{B}, B\Big), 1, L\Big).
\end{equation}

